\documentclass[structabstract]{raa_rb}
\usepackage{soul}
\usepackage{color}
\usepackage{booktabs}
\usepackage{graphicx,times}
\usepackage{natbib,amssymb}
\usepackage{epstopdf,epsfig}
\usepackage{float}
\usepackage{amsmath}
\usepackage{ulem}
\usepackage{amsmath}

\usepackage{mathtools}

\providecommand{\bm}[1]{\mbox{\boldmath$#1$\unboldmath}}

\begin{document}
   \title{On the surface helium abundance of B-type hot subdwarf stars from the WD+MS channel of Type Ia supernovae}

   \volnopage{ {\bf 2024} Vol.\ {\bf X} No. {\bf XX}, 000--000}
   \setcounter{page}{1}

   \author{Rui-Jie Ji 
      \inst{1,2,3,4}
   \and Xiang-Cun Meng
      \inst{1,3,4}
  \and Zheng-Wei Liu
      \inst{1,3,4}
   }

   \institute{Yunnan Observatories, Chinese Academy of Sciences, Kunming 650216,
             Kunming 650216, China; {\it jiruijie@ynao.ac.cn,xiangcunmeng@ynao.ac.cn,zwliu@ynao.ac.cn}\\
        \and
             University of Chinese Academy of Sciences, Beijing 100049, China\\
        \and
             Key Laboratory for the Structure and Evolution of Celestial Objects, Chinese Academy of Sciences,
             Kunming 650216, China\\
         \and
             International Centre of Supernovae (ICESUN), Yunnan Key Laboratory, Kunming 650216, China\\}

   \date{Received ; accepted}

\abstract{ The origin of intermediate helium (He)-rich hot subdwarfs are still unclear. Previous studies have suggested that some surviving Type Ia supernovae (SNe Ia) companions from the white dwarf~+~main-sequence (WD+MS) channel may contribute to the intermediate He-rich hot subdwarfs. However, previous studies ignored the impact of atomic diffusion on the post-explosion evolution of surviving companion stars of SNe Ia, leading to that they could not explain the observed surface He abundance of intermediate He-rich hot subdwarfs. In this work, by taking the atomic diffusion and stellar wind into account, we trace the surviving companions of SNe Ia from the WD+MS channel using the one-dimensional stellar evolution code \textsc{MESA} until they evolve into hot subdwarfs. We find that the surface He-abundances of our surviving companion models during their core He-burning phases are in a range of $-1 \lesssim {\rm log}(N_{\rm He}/N_{\rm H}) \lesssim 0$, which are consistent with those observed in intermediate He-rich hot subdwarfs. This seems to further support that surviving companions of SNe Ia in the WD+MS channel are possible to form some intermediate He-rich hot subdwarfs.
}

\titlerunning{On the surface He abundance of sdBs}

\authorrunning{Ji et al.}

\maketitle
\section{INTRODUCTION} \label{1. INTRODUCTION}

Because of their consistent peak luminosities, type Ia supernovae (SNe Ia) are considered to be one of the most reliable distance indicators. As a result, SNe Ia are utilized to quantify cosmological parameters, which helped researchers reveal the Universe's rapid expansion. This suggests that dark energy is in charge of our universe (\citealt{1998AJ....116.1009R, 1999ApJ...517..565P}). In order to evaluate the dark energy equation and its temporal evolution, SNe Ia are also employed as cosmic probes (\citealt{2011NatCo...2..350H, 2011ApJ...737..102S}).

Even though SNe Ia are crucial to contemporary astrophysics, there is ongoing discussion over a few fundamental issues, including their progenitor systems and explosion mechanism (\citealt{2000ARA&A..38..191H, 2000A&ARv..10..179L, 2012NewAR..56..122W, 2014ARA&A..52..107M,Liu2023RAA}). The majority of researchers agree that a close binary which includes one carbon-oxygen white dwarf (CO WD) is where supernovae (SNe) Ia started (\citealt{2000ARA&A..38..191H,2004Natur.431.1044B, 2011Natur.480..344N, 2013FrPhy...8..116H}). Thus, the progenitor models of SNe Ia are primarily separated into the single-degenerate (SD) model and the double-degenerate (DD) model based on the different sorts of companions (\citealt{1973ApJ...186.1007W, 1982ApJ...253..798N, 1984ApJS...54..335I, 1984ApJ...277..355W}). The non-degenerate companion star in the SD model could be a main-sequence star (i.e. the WD+MS channel), a sub-giant, a red giant star (i.e. the WD+RG channel), an asymptotic giant branch star (i.e. the WD+AGB channel), or a helium star (i.e. the WD+He~star channel) (\citealt{1973ApJ...186.1007W, 1984ApJ...286..644N, 2009MNRAS.395..847W, 2012NewAR..56..122W, 2023RAA....23g5010L}). The binary system in the DD model is made up of two CO WDs (\citealt{1998MNRAS.296.1019H,2018MNRAS.473.5352L}). While no surviving companion is left in the standard DD model, the SD model predicts that the companion stars will survive from the SN Ia explosion. Finding the surviving companions in nearby SN remnants (SNRs) is therefore a promising way to differentiate between the SD and DD models. As a consequence, a detailed analysis of the characteristics of the potential surviving companions is believed to be useful in understanding the origins of SN Ia (\citealt{2017ApJ...836...85L, 2019MNRAS.482.5651M, 2023FrASS..1012880R}).

Hot subdwarfs have been proposed to be possible candidates of surviving companions of SNe Ia \citep[e.g.][]{2019MNRAS.482.5651M}. They are core helium-burning stars with a very thin hydrogen envelope, and they are generally divided into B-type hot subdwarf stars (sdBs) and O-type hot subdwarf stars (sdOs) according to their spectral features (\citealt{1986ApJS...61..305G}). In the Hertzsprung–Russell diagram, hot subdwarfs are either located at the blue end of the horizontal branch or have evolved past that stage (\citealt{2009ARA&A..47..211H}). SdBs typically have effective temperatures between 20,000K and 40,000K, while sdOs have effective temperatures between 40,000K and 80,000K, and are generally more luminous than sdBs. Some mysteries regarding the genesis of hot subdwarfs remain. Given that a significant portion of sdB stars are found in close binaries, processes involving binary interaction should be responsible for their formation (\citealt{2009ARA&A..47..211H, 2016PASP..128h2001H}). 
A detailed description of how hot subdwarfs originate via binary evolution, specifically through the channels of Roche Lobe overflow (RLOF), common-envelope ejection (CE) and double He WDs merger, can be found in \cite{2002MNRAS.336..449H, 2003MNRAS.341..669H}.

Most of hot subdwarfs are observed to present some peculiar chemical abundance signatures. The majority of sdBs have a He deficient atmosphere, and their surface He abundance may be as little as a thousandth of the solar value. From one percent of the solar value to nearly pure He atmosphere, the sdOs display a range of surface He abundances (\citealt{2009ARA&A..47..211H, 2016PASP..128h2001H}).
Previous researchers have categorized hot subdwarfs into He-deficient and He-rich hot subdwarfs based on the solar helium abundance ${\rm log}(N_{\rm He}/N_{\rm H})=-1$, where $N_{\rm H}$ and $N_{\rm He}$ represent the surface number densities of hydrogen and helium, respectively (\citealt{2012MNRAS.427.2180N,2016ApJ...818..202L, 2019ApJ...881....7L}). He-rich hot subdwarfs are further divided into extreme He-rich (eHe-rich) stars (${\rm log}(N_{\rm He}/N_{\rm H})>0$) and intermediate He-rich (iHe-rich) stars ($-1<{\rm log}(N_{\rm He}/N_{\rm H})<0$).
Generally, it is believed that the He-rich group results from the merger of two He WDs, whereas the He-deficient group most likely originates from the RLOF and CE ejection channels (\citealt{2002MNRAS.336..449H, 2003MNRAS.341..669H, 2009ARA&A..47..211H, 2016PASP..128h2001H}). But it's still unclear where the iHe-rich group originated (\citealt{2017MNRAS.467...68M,2019ApJ...881....7L}).

\cite{2017MNRAS.469.4763M} proposed a new version of SD model for SN Ia, i.e. the common-envelope wind (CEW) model. In the model, if the mass transfer rate between a CO WD and its companion exceeds a critical accretion rate, a common envelope forms around the binary system. The WD may gradually increase its mass at the base of the CE.
Based on the CEW model, \cite{2021MNRAS.507.4603M} have studied the formation of hot subdwarfs from the surviving companions of SNe Ia in the WD+MS channel \citep{2019MNRAS.482.5651M,2020ApJ...903..100M}, and then comparing their results with some observational properties of iHe-rich hot subdwarfs. They discovered that several observational features, such as their effective temperatures and surface gravities, may be explained by the hot subdwarfs from this channel. However, their models are still difficult to explain the observed surface He-abundances of iHe-rich subdwarfs. This may be because that the effect of the atomic diffusion and wind mass loss from the surface of the sdB star were ignored in their calculations.

Previous studies have shown that the the atomic diffusion and wind mass loss from the surface of the sdB star could considerably affect their surface He abundances \citep{2001A&A...374..570U,2011MNRAS.418..195H}. Therefore, it will be still important to investigate whether or not the inclusion of the atomic diffusion and wind mass loss during the evolution of surviving companion stars of SNe Ia in the WD+MS channel could explain the observed surface He-abundance of iHe-rich subdwarfs. Therefore, by taking the atomic diffusion and stellar wind into account, we follow the evolution of SNe Ia's surviving companions from the WD+MS channel until they evolve hot subdwarfs using the one-dimensional stellar evolution code MESA in this work. We describe our methods in Section 2. Section 3 displays the results, while Section 4 provides the discussions and summaries.

\section{METHODS} \label{2.METHODS}

\subsection{Atomic diffusion}
The phrase "atomic diffusion" refers to a group of particle transport mechanisms that alter a star's chemical composition. Radiative levitation, thermal diffusion, concentration diffusion, and gravitational settling are some of these mechanisms (\citealt{2010A&A...511A..87H}). The rivalry between outward radiant forces and inward gravity determines an element's actual diffusion velocity, which varies depending on the element (\citealt{2010A&A...511A..87H,2011MNRAS.418..195H}). As such, the abundance profile of a star is directly impacted by atomic diffusion (\citealt{2018MNRAS.475.4728B}). Because the diffusion timescale is roughly proportional to the density of protons, atomic diffusion is more efficient in a star's outer layers (\citealt{2018A&A...618A..10D}). Consequently, atomic diffusion must be included in order to accurately forecast surface abundance, particularly for stars with high surface gravity like hot subdwarfs (\citealt{2022A&A...659A.162C}).

The atomic diffusion in a hot subdwarf from the WD+MS SN Ia channel is studied using the stellar evolution code Modules for Experiments in Stellar Astrophysics (MESA). MESA includes concentration diffusion, thermal diffusion and gravitational settling as the standard processes with the $do\_element\_diffusion$ flag. In MESA, the atomic diffusion is computed by the use of the \cite{1994ApJ...421..828T} and \cite{2011MNRAS.418..195H} formalism to solve the Burgers equations (\citealt{1969fecg.book.....B}). The radiative levitation is included as an optional method (\citealt{2011ApJS..192....3P,2013ApJS..208....4P,2015ApJS..220...15P,2018ApJS..234...34P,2019ApJS..243...10P}). In this paper, we just focus on the surface helium abundance. Generally, the effect of radiative levitation on the helium is not significant, while is remarkable to heavy elements (\citealt{2010A&A...511A..87H,2011MNRAS.418..195H}). Thus, we do not take into account the effect of radiation levitation in this study since calculating radiative levitation in MESA takes a lot of time.

\subsection{Subdwarf models}

\begin{figure}
   \centering
   \includegraphics[width=0.8\textwidth, angle=0]{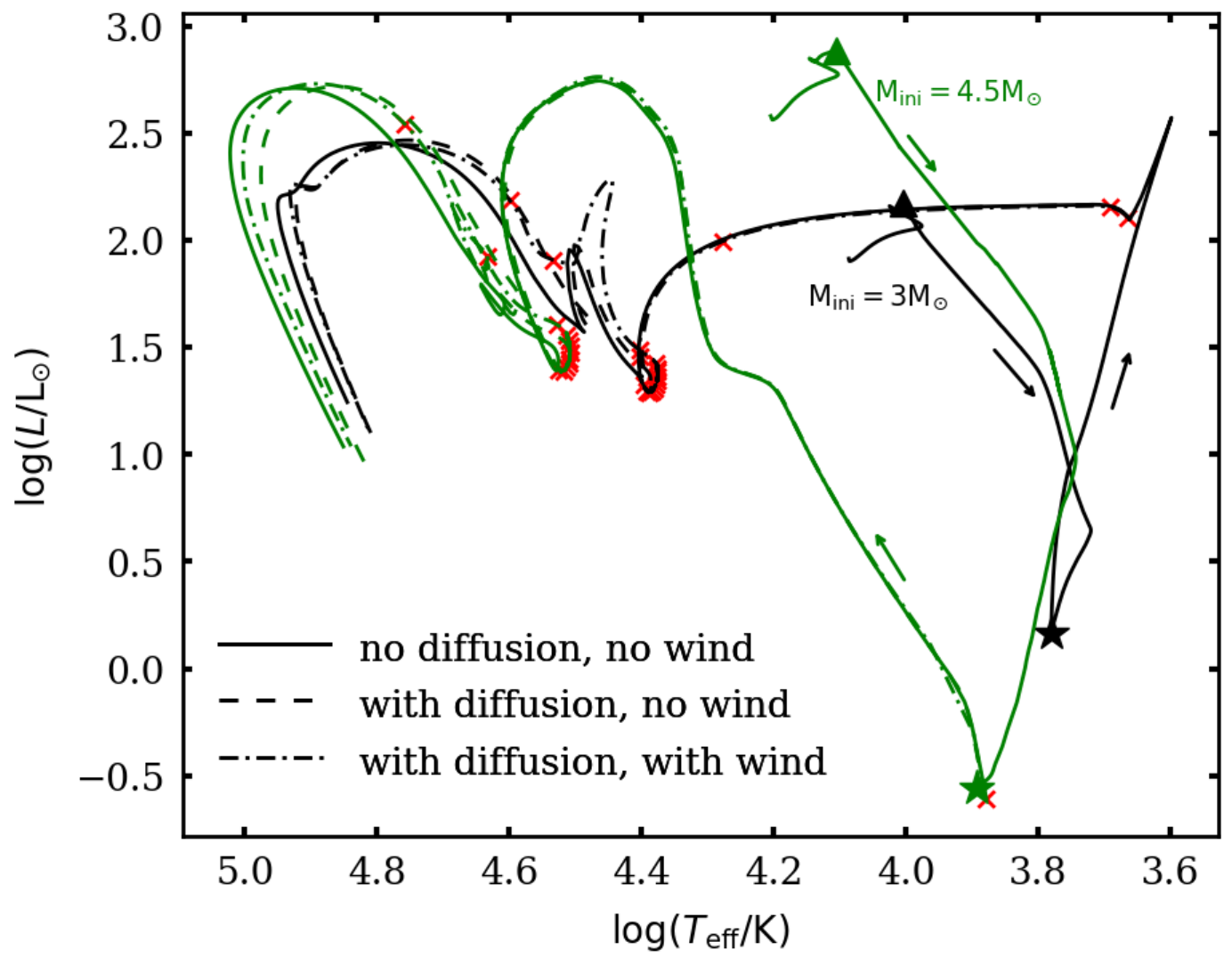}
   \caption{The evolutional tracks of two companion stars with initial mass of 3 $M_\odot$ (black lines, model A) and 4.5 $M_\odot$ (green lines, model B) in the Hertzsprung-Russell (HR) diagram. The triangles represent the position where the companions start to lose mass. The asterisks represent the position where an SN Ia explosion is assumed. The arrows direct the evolutionary direction. The age gap in each line between two consecutive red crosses is $10^7$ years.}
   \label{Fig1}
\end{figure}

In the WD + MS channel, a WD accretes hydrogen-rich materials from the companion star when the companion fills its Roche lobe on the MS or in the Hertzsprung gap (HG) (\citealt{2017MNRAS.469.4763M}). The WD explodes as a SN Ia when its mass over the Chandrasekhar mass limit. The companion may continue to evolve into a sdB star. Since we are only concerned with the surface helium abundance of the companion star during the sdB phase, we use single-star evolution instead of binary evolution calculation to construct sdB models. The mass transfer of the companion is then modeled with a constant mass-loss rate.

For systems with different initial masses, we display the evolution of the models in the Hertzsprung–Russell (HR) diagram in Fig.~\ref{Fig1}. First, we choose two MS stars of 3 $M_\odot$ (model A) and 4.5 $M_\odot$ (model B) with solar abundances (X = 0.70, Y = 0.28, Z = 0.02), and evolve them to the stage of the HG phase. Then, we activate the high mass-loss rate option (the $mass\_change$ option in MESA) to simulate the mass loss of the star. For model A, the mass loss timescale is $2.5 \times 10^6$ years, and the mass loss rate is $10^{-6}M_\odot/{\rm yr}$. For model B, the mass loss timescale is $3.98 \times 10^5$ years, and the mass loss rate is $10^{-5}M_\odot/{\rm yr}$. Then, we assume a SN Ia occurs and the mass loss stops. The systems with more massive WD, and more massive companion, the more likely to form hot subdwarfs, which means a higher mass loss rate and then a shorter mass loss rate. The employed mass loss timescale and mass loss rate used here are comparable to those of typical binary evolutions for SNe Ia (\citealt{2017MNRAS.469.4763M}), and the hot subdwarfs stars from such a treatment are similar to those from a system of (${\rm log}P^{\rm i}/{\rm d}, M^{\rm i}_{\rm WD}/{\rm M}_{\odot}$) = (0.5, 1.1) (\citealt{2021MNRAS.507.4603M}). During the mass loss phase, the companion stars lose almost all of their hydrogen envelope, which makes the diffusion effect insignificant. Then, before the SN explosion, we do not consider the atomic diffusion process in our models.

During the following evolution, we carry out three cases with different physical inputs. Model A0/B0 do not include the atomic diffusion and the stellar wind, i.e. similar to that in \cite{2021MNRAS.507.4603M}. Model A1/B1 only include the atomic diffusion, and model A2/B2 include both the atomic diffusion and the stellar wind. We use the Reimers' wind to simulate the wind mass loss from the surface of the surviving companion, where $\eta_{\rm Reimers}$ is set to 0.1 (\citealt{1975MSRSL...8..369R}).

\section{RESULTS} \label{3. RESULTS}

\subsection{Evolution to the sdB}

After the rapid mass loss phase, the masses of models A and B are $0.5 {\rm M}_\odot$ and $0.52 {\rm M}_\odot$, respectively, and the hydrogen envelope masses are $0.2 {\rm M}_\odot$ and $0.02 {\rm M}_\odot$, respectively.
The following evolutions of the stars are also shown in Fig.~\ref{Fig1}. For model A, the star consecutively experience the red giant branch (RGB) and the horizontal branch (HB) phases. Its hydrogen envelope is consumed by shell hydrogen burning during the HB phase, causing the envelope to grow gradually thinner. For the consumption of the envelope, its surface temperature becomes higher and higher, and then, the star evolve to the hot subdwarf stage in HR diagram. After the exhaustion of the helium in the core, the subdwarf directly evolves to the WD branch, rather than to the asymptotic giant branch (AGB). For model B, the star will expand rapidly, and then enter the sdB stage about two million years after mass loss is stopped (see also in \citealt{2020ApJ...903..100M}).
Due to its larger mass, model B has a higher surface effective temperature compared to model A during the sdB phase. Its subsequent evolution track is similar to model A. The differing post-SN Ia evolutions of models A and B may be traced back to the varying masses of the hydrogen-rich envelope when the supernova explosion occurs; in other words, the evolution track of a star with thicker hydrogen envelope is more
similar to that of an isolated star. 

\begin{figure}
   \centering
   \includegraphics[width=0.8\textwidth, angle=0]{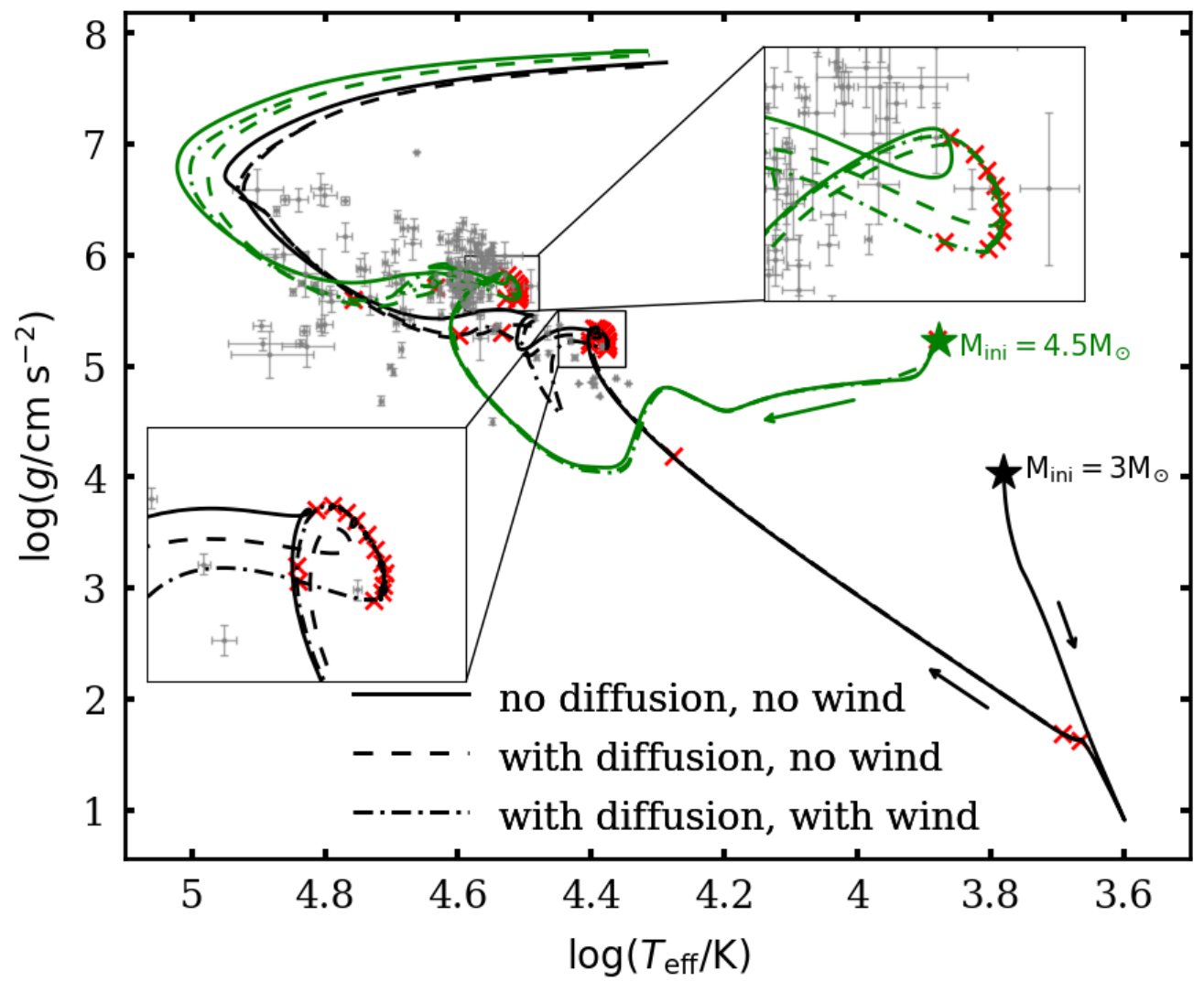}
   \caption{The evolutionary tracks of two companion stars with initial mass of 3 $M_\odot$ (black lines, model A) and 4.5 $M_\odot$ (green lines, model B) in a ${\rm log}\,g - {\rm log}\,T_{\rm eff}$ diagram. The asterisks represent the position where an SN Ia explosion is assumed. The arrows represent the evolutionary direction. The age gap in each line between two consecutive red crosses is $10^7$ years. The grey dots belongs to iHe-rich group, and the data are from \cite{2018ApJ...868...70L, 2019ApJ...881..135L, 2020ApJ...889..117L} and \cite{2019ApJ...881....7L, 2021ApJS..256...28L}.}
   \label{Fig2}
\end{figure}

We present the evolution of the companions in the ${\rm log}g-{\rm log}T_{\rm eff}$ diagram in Fig.~\ref{Fig2} in order to compare with observations, where some iHe-rich subdwarfs are also shown and the observational data are from \cite{2018ApJ...868...70L, 2019ApJ...881..135L, 2020ApJ...889..117L} and \cite{2019ApJ...881....7L, 2021ApJS..256...28L}. The iHe-rich hot subdwarf samples shown in here are isolated stars. As shown in Fig.~\ref{Fig1}, different physical inputs cannot affect the evolutions of the surviving companions in ${\rm log}\,g - {\rm log}\,T_{\rm eff}$ diagram. Fig.~\ref{Fig2} shows that the evolutionary tracks span several regions of iHe-rich hot subdwarfs, i.e. the surviving companion model may explain several observational properties of the iHe-rich subdwarfs as suggested by \cite{2021MNRAS.507.4603M}. In fact, the initial binary parameters of the systems generating SNe Ia largely determine the features of hot subdwarfs from the WD+MS channel. \cite{2021MNRAS.507.4603M} studied many hot subdwarfs models from WD+MS channel with different initial mass and orbit period. If the initial orbital period is longer (mass transfer occurs relatively later) and the initial mass of the companion star is larger (the surviving companion may be relatively more massive), the companion may have a higher effective temperature and a higher surface gravity in the hot subdwarf phase. Additionally, we notice that the evolution track of model A1/B1 (dashed line) is somewhat pushed to lower surface gravities and effective temperatures when compared to model A0/B0 (solid line). This is because of the outward diffusion of H, which is followed by a drop in envelope density. As a result, the sdB star becomes somewhat bigger and colder, and the envelope becomes less gravitationally bound. However, the difference is so insignificant that can not be discriminated by observations. Therefore, the location of the hot subdwarf stars in the HR or ${\rm log}g-{\rm log}T_{\rm eff}$ diagrams is not much affected by atomic diffusion, at least for those from the SNe Ia channel.

\subsection{Surface helium abundance}

\begin{figure}
   \centering
   \includegraphics[width=0.9\textwidth, angle=0]{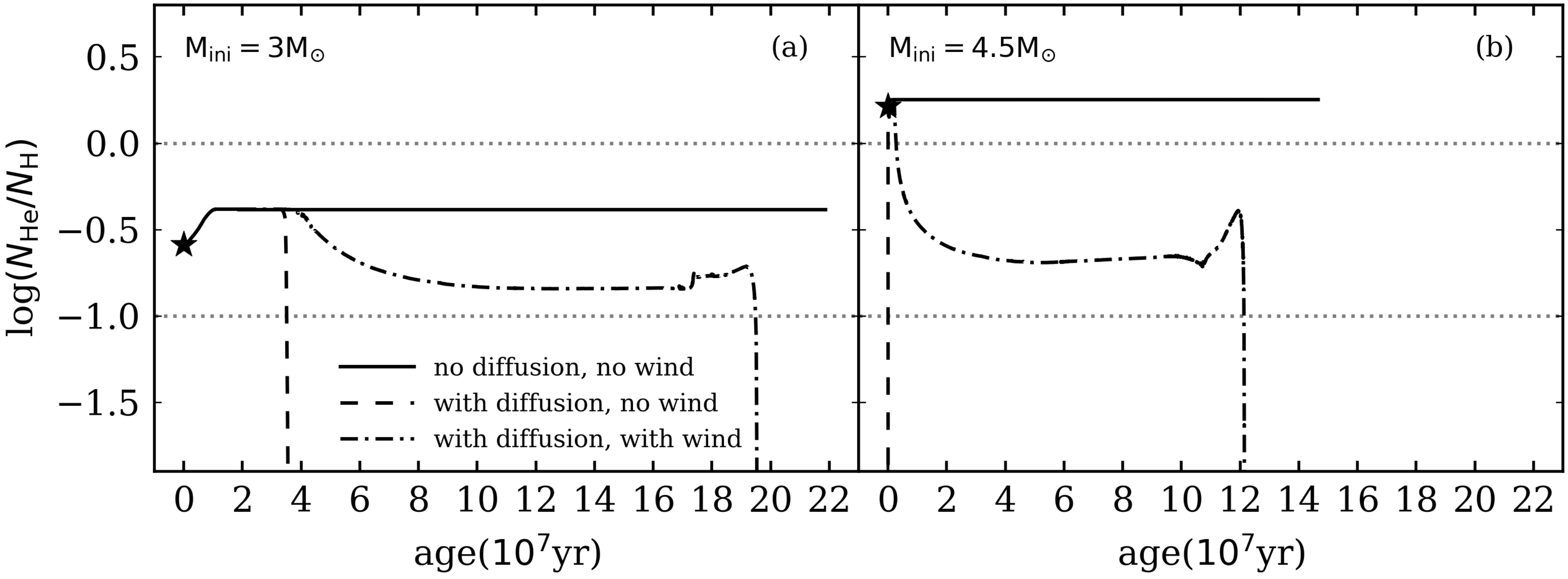}
   \caption{The evolution of ${\rm log}(N_{\rm He}/N_{\rm H})$ with time, where the surface number densities of hydrogen and helium are denoted by $N_{\rm H}$ and $N_{\rm He}$, respectively. The range of He abundance for iHe-rich hot subdwarfs is shown by the two dotted lines. The asterisks represent the position where an SN Ia explosion is assumed.}
   \label{Fig3}
\end{figure}

After the supernova explosions, the surface helium abundance for the model A and B are about ${\rm log}(N_{\rm He}/N_{\rm H}) = -0.59$ and ${\rm log}(N_{\rm He}/N_{\rm H}) = 0.24$, respectively. The following evolution of ${\rm log}(N_{\rm He}/N_{\rm H})$ depends on the physical inputs and the evolutionary stage the stars experiencing. In Fig.~\ref{Fig3}, we show the evolutions of ${\rm log}(N_{\rm He}/N_{\rm H})$ for model A (panel a) and model B (panel b), respectively, where the surface number densities of hydrogen and helium are denoted by $N_{\rm H}$ and $N_{\rm He}$, respectively. In panel (a), there is an increase in surface helium abundance for model A due to the first dredge up in the RG phase, where the atomic diffusion does not take effect until the surface convection ceases. Model B does not experience the RG stage, and then, the atomic diffusion takes effect immediately, as shown in panel (b) of Fig.~\ref{Fig3}. For model A0/B0 (solid line), the surface helium and hydrogen abundance are not changed during the whole subdwarf stage and the surface helium abundance of model A0 is consistent with the iHe-rich hot subdwarfs from observation. For model A1/B1 (dashed line), as shown in previous studies (\citealt{1982ApJ...254..221W, 1997fbs..conf..169F}), the helium on the surface quickly settles down below the photosphere within about $10^4$ years for the effect of gravitational settling, and the surface almost becomes a pure hydrogen atmosphere, i.e. ${\rm log}(N_{\rm He}/N_{\rm H})$ becomes lower than $10^{-15}$. Then in Fig.~\ref{Fig3}, we do not show the whole range of ${\rm log}(N_{\rm He}/N_{\rm H})$ for model A1/B1. For model A2/B2 (dot-dashed line), the helium will also settle down as model A1/B1. However, ${\rm log}(N_{\rm He}/N_{\rm H})$ will keep at a level between 0 and -1 for about $1.8 \times 10^8$years for model A2 and $1 \times 10^8$years for model B2 due to the presence of stellar wind, until the stars evolve to the WD branch. In other words, the stars behave as iHe-rich subdwarfs. 

\begin{figure}
   \centering
   \includegraphics[width=0.8\textwidth, angle=0]{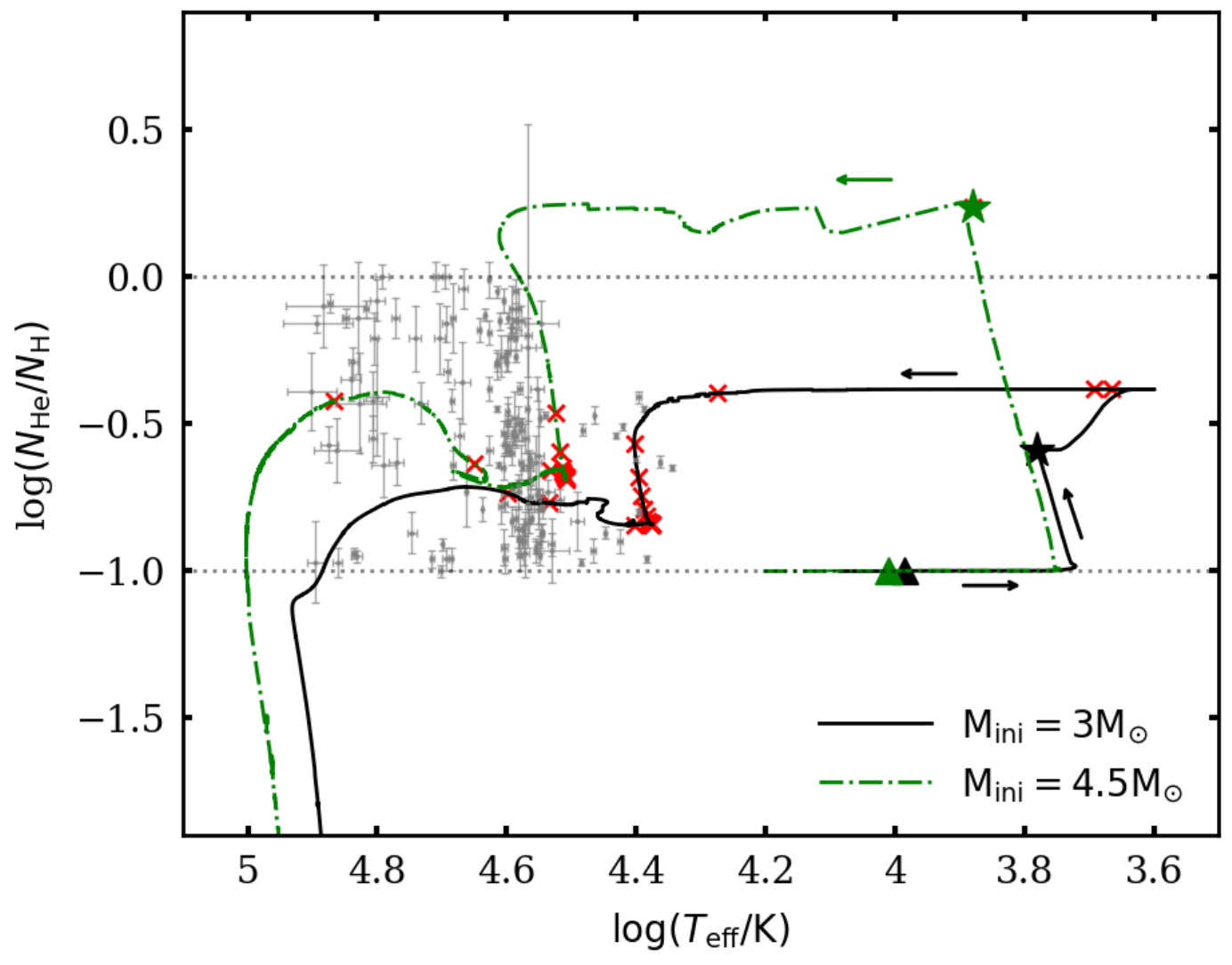}
   \caption{The evolution of ${\rm log}\,(T_{\rm eff})$ and ${\rm log}(N_{\rm He}/N_{\rm H})$ of models with diffusion and wind. The arrows represent the evolutionary direction, and the asterisks represent the position where an SN Ia explosion is assumed. The age gap in each line between two consecutive red crosses is $10^7$ years. The range of He abundance for iHe-rich hot subdwarfs is shown by the two dotted lines. The grey dots belongs to iHe-rich group, and the data are from
   \cite{2018ApJ...868...70L, 2019ApJ...881..135L, 2020ApJ...889..117L} and \cite{2019ApJ...881....7L, 2021ApJS..256...28L}.}.
   \label{Fig4}
\end{figure}

Actually, the surface helium abundance during the hot subdwarfs heavily depends on remaining the mass of the hydrogen-rich envelope after the SN explosion. Usually, the less massive the envelope, the higher the surface helium abundance in the sdB phase. To compare with the observations of iHe-rich hot subdwarfs, we demonstrate the evolution of model A2/B2 in the ${\rm log}\,(T_{\rm eff}) - {\rm log}(N_{\rm He}/N_{\rm H})$ diagram in Fig.~\ref{Fig4}. This figure represents the companions following the supernova explosion spend most of their time in the area of iHe-rich hot subdwarfs, if the effect of atomic diffusion and stellar wind are considered. Again, the parameter of the initial model has a great influence on the position in ${\rm log}\,(T_{\rm eff}) - {\rm log}(N_{\rm He}/N_{\rm H})$ plane for our models.

\subsection{Helium abundance profile}

\begin{figure}
   \centering
   \includegraphics[width=0.9\textwidth, angle=0]{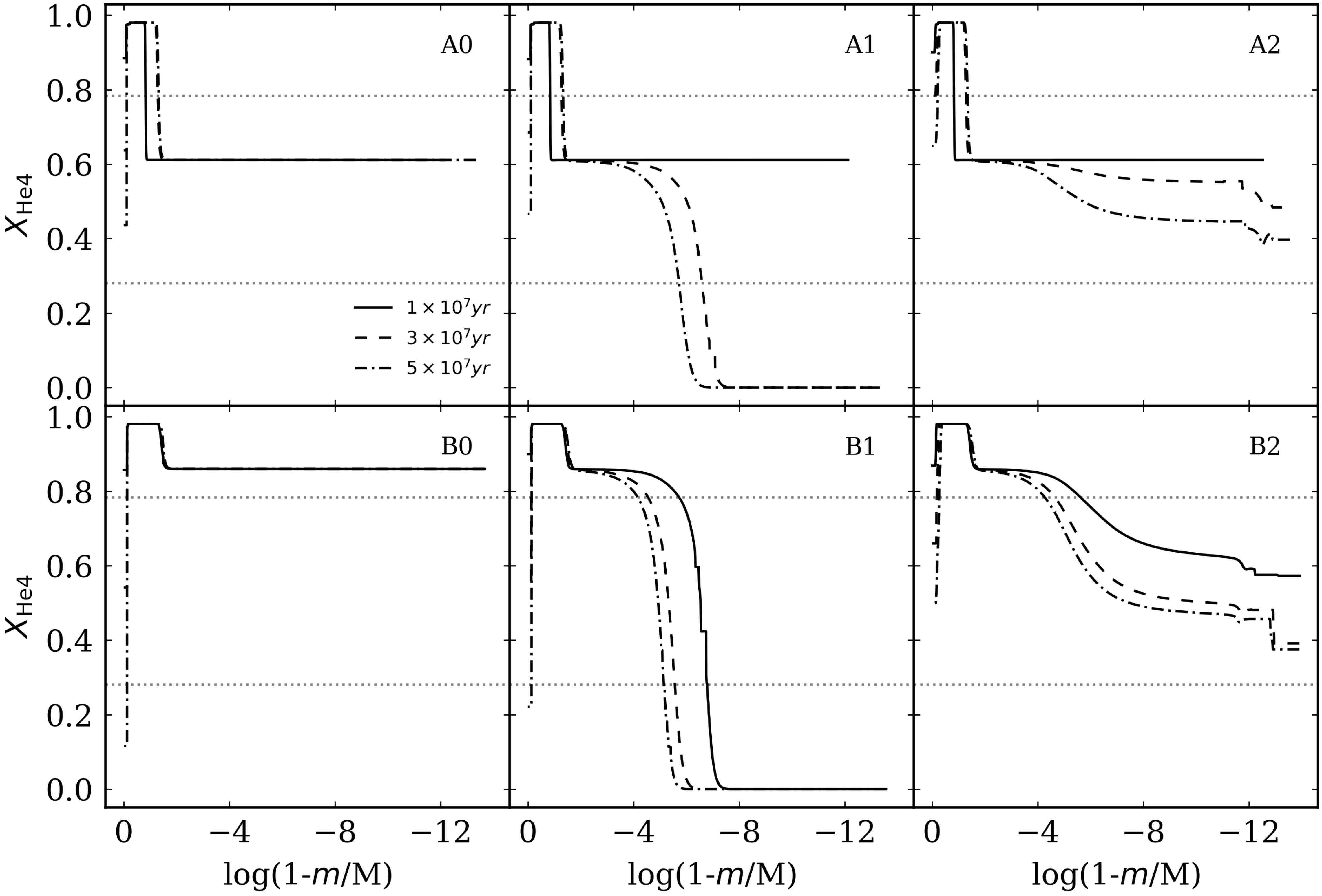}
   \caption{Snapshots of the helium abundance profile for different models. Different lines represent different time points after the SN Ia explosion. The range of He abundance for iHe-rich hot subdwarfs is shown by the two dotted lines.}
   \label{Fig5}
\end{figure}

The He-profile versus the fractional mass ${\rm log}(1-m/M)$ at three different ages of the companion stars — $1 \times 10^7$, $3 \times 10^7$, and $5 \times 10^7$ yr after the supernova explosion — is shown in Fig.~\ref{Fig5} to demonstrate the influence of a time-dependent diffusion. Without the atomic diffusion, the helium abundance does not change in the envelope, as shown in the panel A0/B0 of Fig.~\ref{Fig5}. The outward movement of the helium core is due to the burning of the hydrogen shell. If we consider the diffusion only, the surface helium settles down quickly and the outer layer consists of pure hydrogen. In particular, the outer hydrogen envelope becomes thicker and thicker as shown in panel A1/B1. If both atomic diffusion and stellar wind are considered, although the surface helium will settle down, the stellar wind will simultaneously peel off the outer layer, and then the surface helium abundance can remain at relatively high level, as shown in panel A2/B2. The figure also makes it very clear that the interior profile is seldom impacted by atomic diffusion, and that atomic diffusion works only well in the outer layer, which makes up around 1\% mass of the star.

\section{DISCUSSION AND SUMMARY} \label{4. DISCUSSION AND SUMMARY}

Following \cite{2021MNRAS.507.4603M}, we used MESA to compute different hot subdwarf models with or without the effects of the atomic diffusion and the stellar wind. For the basic models with no diffusion, if the surface convention ceases, surface helium abundance remains constant after supernova explosion. With the atomic diffusion included, surface helium are depleted quickly when the surface convection ceases. However, if a stellar wind is also included, the surface helium abundance can keep at the iHe-rich level during the whole sdB phase.

To simplify the calculation, we used single stars evolution to replace binary evolution in this paper. Therefore, we have ignored some interactions in binary evolution, which may have some impact on our results. For example, in binary evolution, the ejecta from a supernova explosion can strip away portions of the envelope of the companion star (\citealt{2000ApJS..128..615M,2007PASJ...59..835M,2012A&A...548A...2L,2010ApJ...715...78P,2019ApJ...887...68B,2022MNRAS.514.4078M}), which might lead to a slight increase in the surface helium abundance of the companion (see also the discussion in \cite{2020ApJ...903..100M}). In this paper, we just evolve two models, because the calculation of the atomic diffusion is a very time-consuming task in MESA. However, \cite{2021MNRAS.507.4603M} studied many star models from WD+MS channel, and different initial masses and orbit periods are taken into account in detail. In the ${\rm log}\,g - {\rm log}\,T_{\rm eff}$ diagram, their models could cover most samples of the iHe-rich hot subdwarfs. The basic conclusions based on the two model may not be changed by different initial models. In addition, we used the Remiers' wind as the mass-loss rate in our model. The physics of the wind of hot subdwarfs may be complex (\citealt{2016A&A...593A.101K}). We did some tests with stellar wind of different constant mass loss rates which is similar to test different stellar wind coefficients, and found that the weaker the stellar wind, the lower the surface helium abundance of our model in the hot subdwarf phase. Thus, different stellar wind models and wind coefficients could produce hot subdwarfs with different surface abundances, which need to be studied in more detail in the future. Moreover, the abundance of surface elements is not only affected by the stellar wind and the atomic diffusion, but also affected by other physical mechanisms, such as surface rotation, magnetic field, turbulent mixing, etc. For example, in the work of \cite{2011MNRAS.418..195H}, they studied a typical hot subdwarf of $0.46 M_{\odot}$, with mixing of the outer $\Delta M \approx 10^{-8},10^{-7}$ and $10^{-6} M_{\odot}$, respectively. Their models reproduced the observed He abundances of He-deficient ones.

Though a few of evolutionary situation about iHe-rich hot subdwarfs have been proposed, including the merging of a low-mass MS star star with a He WD (\citealt{2017ApJ...835..242Z}), a post-common envelope system with incomplete atmosphere stratification (\citealt{2012MNRAS.423.3031N}), the late hot-flasher scenario (\citealt{2008A&A...491..253M}), the formation of iHe-rich ones remains unclear. As described in \cite{2021MNRAS.507.4603M}, the atmosphere of the companion may be polluted by supernova ejecta. As a result, the iron-peak elements could be enhanced in the surface of the hot subdwarfs from the supernova channel. It must be noted that the SN Ia channel provides just part of the iHe-rich hot subdwarfs with specific Galactic kinematic features, so it is not possible to be the main contributor. Thus, the origin of iHe-rich hot subdwarfs and the physical mechanisms affecting the helium abundance on the surface of hot subdwarfs remain to be investigated in detail.

\begin{acknowledgements}
We thank the referees for their valuable comments. This study is supported by the National Natural Science Foundation of China (Nos. 12288102 and 12333008) and the  National Key R\&D Program of China (No. 2021YFA1600403). X.M. acknowledges support from the Yunnan Ten Thousand Talents Plan - Young \& Elite Talents Project, and the CAS ‘Light of West China’ Program. X.M. acknowledges support from International Centre of Supernovae, Yunnan Key Laboratory (No. 202302AN360001), the Yunnan Revitalization Talent Support Program-Science \& Technology Champion Project (NO. 202305AB350003), Yunnan Fundamental Research Projects (NOs. 202401BC070007, 202201BC070003) and the science research grants from the China Manned Space Project.
\end{acknowledgements}

\bibliographystyle{raa}
\bibliography{main}

\end{document}